# EfficientNet for Brain-Lesion classification

Quoc-Huy Trinh[1], Trong-Hieu Nguyen Mau[1], Radmir Zosimov[2], and Minh-Van Nguyen[1]

[1] Faculty of Information Technology, University of Science, VNU-HCM, Ho Chi Minh city, Vietnam
[2] SBEI School №1228 "Lefortovo"
{20120013, 20120081, 20127094}@student.hcmus.edu.vn, radmir.zosimov@gmail.com

**Abstract.** In the development of technology, there are increasing cases of brain disease, there are more treatments proposed and achieved a positive result. However, with Brain-Lesion, the early diagnoses can improve the possibility for successful treatment and can help patients recuperate better. From this reason, Brain-Lesion is one of the controversial topics in medical images analysis nowadays. With the improvement of the architecture, there is a variety of methods that are proposed and achieve competitive scores. In this paper, we proposed a technique that uses efficient-net for 3D images, especially the Efficient-net B0 for BrainLesion classification task solution, and achieve the competitive score. Moreover, we also proposed the method to use Multiscale-EfficientNet to classify the slices of the MRI data.

## 1 Introduction

In recent years, the number of cases that have brain lesions increasing, according to the National Brain Tumor Society, in the United States, about 700,000 people live with a brain tumour, and the figure rises by the end of 2020 [20]. Compared with other cancers such as breast cancer or lung cancer, a brain tumour is not more common, but it is the tenth leading cause of death worldwide [17]. According to United States statistics, An estimated 18,020 adults will die this year from brain cancer. Moreover, the brain lesion can have a detrimental impact on the brain of the patients and can make sequelae for the patients on the others organs or their brain. Nowadays, there are various methods to diagnose disease through medical images such as CT-scan, magnetic resonance imaging (MRI), and X-ray.
A brain lesion is the abnormal sympathy of a brain seen on a brain-imaging test, such as magnetic resonance (MRI) or computerized tomography (CT). Brain lesions appear as spots that are different from other tissues in the brain [18]. By this method, the MRI can visualize the abnormal on the slide of the brain[19]. The goal of the 3D-CT scans images classification task is to evaluate various methods to classify the brain lesions in the medical images correctly and efficiently [21]. Parallel to the development of Computer Vision, particularly the Deep Neural Network and Vision Transformer, multiple methods were proposed to classify the abnormal tissue in the organ through the images such as CT scans and MRIs. In recent years, significant advancement has been made in medical science as the Medical Image processing technique, which helps doctors diagnose the disease earlier and easier. Before that, the process is tedious and time-consuming. To deal with this issue, it is necessary to apply computer-aided technology because Medical Field needs efficient and reliable techniques to diagnose life-threatening diseases like cancer, which is the leading cause of mortality globally for patients [5].

In this paper, we propose a method that uses 3D EfficientNet to classify MRI images, with a new approach to using EfficientNet with Multiscale layers (MSL) to classify slices of MRI images. With the 3D EfficientNet, the model can have higher performance on feature extraction and classification task. In contrast, MSL uses the feature on the slice of image and create low-quality features to create a better feature map for the classification task. In this experiment, we use the backbones of EfficientNet B0 and EfficientNet B7 to perform an experiment and evaluation of our method.

## 2    Related work

### 2.1    Image Classification

Image classification is a task that attempts to classify the image by a specific label. The input of the problem is the image the output is the label of this image. In recent years, the development of computing resources leads to a variety of methods in Image classification such as VGG 16, ResNet 50, and DenseNet. These architectures get the competitive result in the specific dataset. With the images sequence dataset, from the previous methods, there are various methods of Convolution Neural Network (CNN) combined with RNN or LSTM have been proposed. In a few years nearby, some Vision Transformer methods, State of the art (SOTA) architecture combined with CNN and CNN 3D have been proposed. These architecture achieve the competitive result on the task they are applied with the performance also has a competitive response on the task they are applied [2].

### 2.2    Transfer Learning

Transfer Learning is the method that applies the previously trained model on the large dataset we can not get access to on the new dataset. The merit of this method is we can use the previous model that has high performance to apply on feature extraction of our dataset, this is the reason why the model with the transfer learning method can achieve better accuracy while training with the small dataset [25].

## 2.3 Brain-Tumour Classification

Brain-Tumour classification is one the most popular tasks in medical image preprocessing [8]. The main goal of this task is to classify the brain lesions images in the set of images. With MRI images, the brain lesion is demonstrated in the dark or light spots, which are different from the others [23].

There are many methods such as segmentation model to improve the data inputs or Generative adversarial networks to increase the data numbers to improve the performance of the training process[22]. Moreover, in recent years, many network architectures have been proposed to improve the classification score of the task [6].

# 3 Dataset

The dataset for the experiment is from BraTS 2021, the target of the dataset is for the brain lesions classification task [14] which is from RSNA-ASNR-MICCAI BraTS 2021 challenge [4]. This dataset consists of 585 cases for training, in each case includes structural multi-parametric MRI (mpMRI) scans and is formatted in DICOM. The exact mpMRI scans included four types are:

- Fluid Attenuated Inversion Recovery (FLAIR)

- T1-weighted pre-contrast (T1w)

- T1-weighted post-contrast (T1Gd)

- T2-weighted (T2) This dataset is seperated in two labels are 0 and 1 for the NGMT value, which is the diagnosis scale of Brain-Tumour Detection.[15]

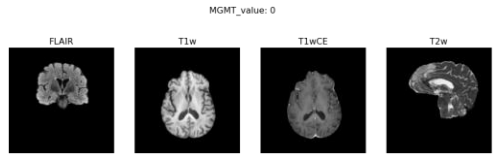

**Fig.1.** Sample of NGMT value 0

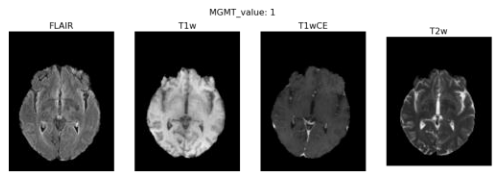

**Fig.2.** Sample of Sample of NGMT value 1

Regarding NGMT promoter methylation status data is defined as a binary label with 0 as unmethylated and 1 is for methylated [16]. In the challenge, this data is provided to the participants as a comma-separated value (.csv) file with the corresponding pseudo-identifiers of the mpMRI volumes.[17] (studylevel label)

## 4    Method

The method we propose in this paper is the classification method for the Brain MRI images data. The input is the Brain MRI Image data (in png, jpg or Dicom format). Then all images will be preprocessed and will be augmented before being trained by the 3D EfficientNet model. Then the model can be used to predict the NGMT value of the new Brain MRI Image data Following is the diagram of our method:

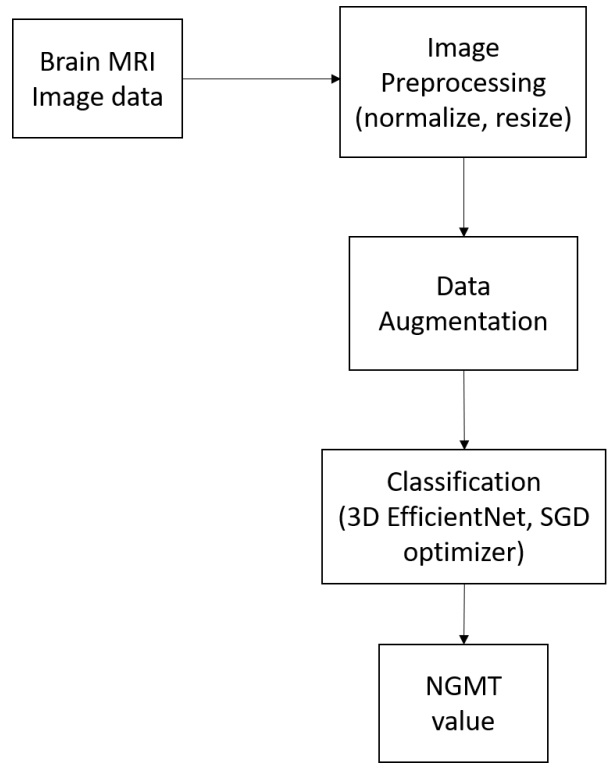

**Fig.3.** Method diagram

With the 2D dataset, we create the data from slices of MRI images depend on 4 index: Flair, T1w, T1Gd and T2. These 4 index can be created to four dataset with different size for each dataset. By using CNN for the 2D images, we can ensemble and probing four data by the ratio 3:3:3:2 and 2:4:2:2 for the result of the experiment.

### 4.1  Data Preparation

After loading data, we resize all the images to the size (256,256), then we split the dataset into the training set and validation set in the ratio of 0.75:0.25. After resizing and splitting the validation set, we rescale the data pixel down in the range [0,1] by dividing by 255, in the MRI data, we can apply rescale data on the slices of the data, as the result, the scale of the data will in the range [0,1]. Then we use the application of EfficientNet to preprocess the input. The input after preprocess is rescaled to the same input of the EfficientNet model.

### 4.2  Data Augmentation

Data Augmentation is vital in the data preparation process. Data Augmentation improves the number of data by adding slightly modified copies of already existing data or newly created synthetic data from existing data to decrease the probability of the Overfitting problem. We use augmentation to generate the data randomly by random flip images and random rotation with an index of 0.2.
With the 2D slices, the augmentation apply on each slices of the MRI data

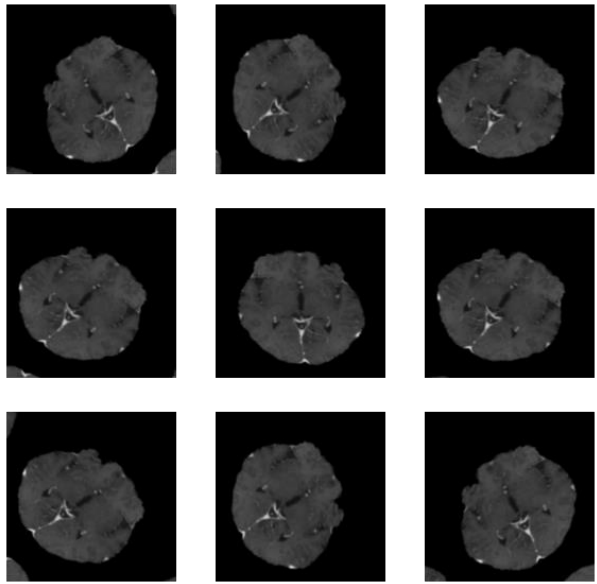

**Fig.4.** The result after data augmentation process

### 4.3 EfficientNet 3D

EfficientNet 3D is the architecture that bases on state-of-the-art 2D EfficientNet architecture. This architecture usually is used for video classification tasks or 3D classification tasks [24]. This architecture has five main parts: Initial Convolutional Layer 3D, Mobile Inverted Residual Bottleneck Block 3D, Convolutional Layer 3D, Global Average Pooling and Fully Connected Layer. This architecture is the modified version for the architecture that uses ConvLSTM or traditional Conv3D layers and it gets competitive scores on the 3D dataset and video dataset[1]. In the experiment, we propose the method by using the input MRI images with the size 256x256x4 to the input of the architecture, after passing through Convolution layer 3D, Mobile Inverted Residual Bottleneck Block 3D, and the others Convolutional layer 3D for the feature extraction, then Global Average Pooling layer will create the feature vector for the classification process.

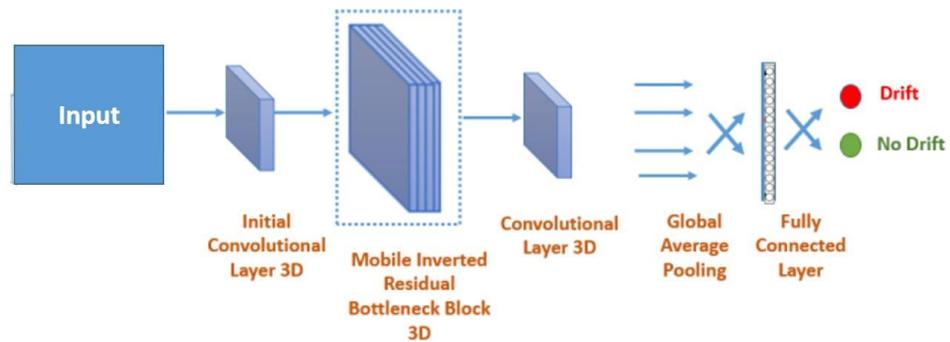

**Fig.5.** EfficientNet3D B0 architecture

### 4.4 Multiscale EfficientNet

In the experiment, we explore that the drawback of using 3D-CNN is the mismatch of the information between the channel space. We approach a new method that uses the slices of the MRI, which are T1-weighted pre-contrast slices. However, the number of slices is adequate for the training process to achieve the well-performance, we propose to use Multiscale block to create the high-quality feature and low-quality feature to ensemble the quality of the feature, then this feature concatenate with the EfficientNet block for the output of the architecture.

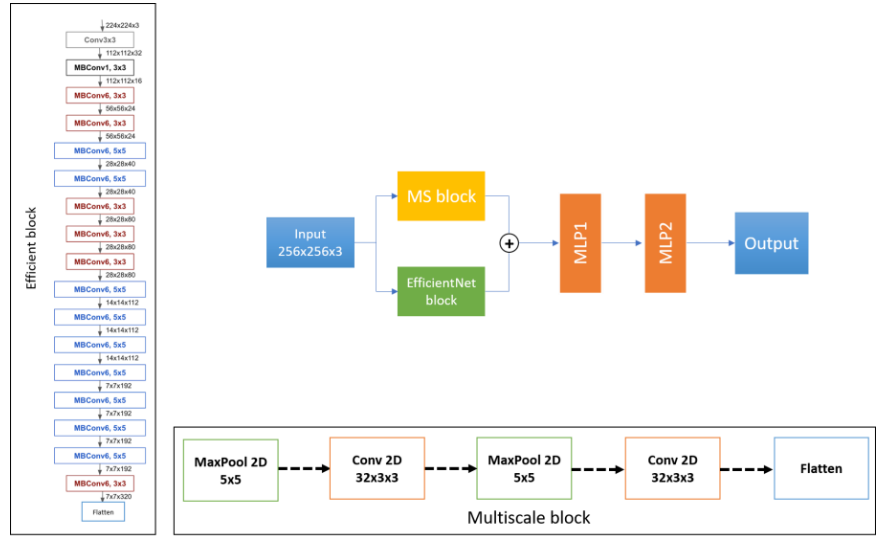

**Fig.6.** Multiscale EfficientNet architecture

From the Input layers with the shape 256x256x3, there are two ways for the input is the Multiscale Block and EfficientNet Block.
We use a Multiscale block containing two Max Pooling layers with two Convolution 2D layers for creating the low-quality feature and for the feature extraction of this feature.
This feature has an integral part of the ensemble and carries more features from the first layers of the MRI slices. From this feature, when add with the highquality feature, the model can get better performance on feature extraction. With the EfficientNet block, the high-quality feature is extracted as the traditional CNN, then the feature output of this block concatenates with the feature of Multiscale block to create the vector with shape output for the classification process.

### 4.5  Loss Function

To evaluate the performance of the model on the training process, we propose to use binary cross-entropy to judge the performance of the model.

$$H_p(q) = -\frac{1}{n}\sum_{i=1}^{N} y_i \cdot \log(p(y_i)) + (1 - y_i) \cdot \log(1 - p(y_i)) \tag{1}$$

Above is the formula for Binary-cross entropy, it is suitable for our binary classification problem.

### 4.6 Optimizer

To get the global minimum in the training process. We do various experiments with optimization such as Stochastic Gradient Descent [12], Adam[9] and Adadelta optimizer[11]. After these experiments, we decide to choose the Adam optimizer because of the merit of the Adam optimizer and the performance of this optimizer on learning rate 0.0001 and the decreasing slightly of validation loss.
Below is the updating formula each weight for Adam optimizer:

$$w_t = w_{t-1} - n\frac{m_t}{\sqrt{v_t} + \epsilon} \tag{2}$$

With adam optimizer, the weight will be updated by the average of the square of the previous slope and it also keeps the speed of slope in the previous as momentum[9]

## 5  Evaluation Metrics

The evaluation of the experiment is demonstrated through an area under the ROC curve (AUC), this is the scale to evaluate the binary classification. For a predictor f, an unbiased estimator of its AUC can be expressed by the WillcoxonMann-Whitney statistic [7]:

$$AUC(f) = \frac{\sum_{t_0 \epsilon D^0} \sum_{t_1 \epsilon D^1} 1[f(t_0) < f(t_1)]}{|D^0| \cdot |D^1|} \tag{3}$$

In this way, it is possible to calculate the AUC by using an average of a number of trapezoidal approximations, it can help to improve the fair in the evaluation phase.

## 6  Evaluation

The following parameters are setup for in this experiment:

Table 1. The parameter setup for model training

| Parameter | Value |
|---|---|
| Optimizer | Adam |
| Learning rate | 0.0001 |
| Backbone | EfficientNet B0 |
| Loss | Binary Crossentropy |
| metrics | AUC |

In the competition, we get an AUC score of 0.60253 on the Test dataset with 87 cases, which is a competitive score. Our methods get a competitive result when compare with the other methods on the same dataset. Below is our experimental evaluation with different optimizers with EfficientNet 3D:

Table 2. The evaluation on each optimizer

| Optimizer | Evaluation |
|---|---|
| Adam | 0.60253 |
| Adadelta | 0.60124 |
| SGD | 0.60178 |
| RMSPROP | 0.60223 |

These evaluations are saved on 100 epochs with the best weight which is evaluated on the validation AUC metrics. After that, we use the Early Stopping method to improve the AUC score of the model by optimizing the calculation of gradient in the optimizer.

For comparison between two approaches and methods, we benchmark two methods with the same test dataset from the organizer.

Table 3. Benchmarking for two methods

| Method | AUC |
|---|---|
| EfficientNet 3D | 0.60253 |
| **Multiscale EfficientNet B7** | **0.67124** |

From the benchmarking table, it is obvious that the performance of the Multiscale EfficientNet B7 is better than the performance of EfficientNet 3D in AUC. However, there are some drawbacks to this method in computing resources. Because creating two types of features are low-quality and high-quality features, the time for computing increases for this process, this is the drawback of this method for running on the lack of computing resources.

## 7  Conclusion

We demonstrated the proposal of using EfficientNet 3D to classify endoscopic images. The result of our research is competitive on the AUC evaluation metric. In our method, we use EfficientNet 3D with Adam optimizer and Early stopping method to improve the performance of the model on the training process to achieve the competitive score. Moreover, we also apply data augmentation to reduce the overfitting problem of the model on the test dataset. However, there are some drawbacks that we have to do to improve the performance of the model, such as pre-processing data, reducing the noise of the training dataset.

Furthermore, we can apply the better backbone of EfficientNet 3D, or we can use the approach of Transformer or spatial Attention modules to have a new approach to per frame of the sequence images. This new approach can get better feature extraction and better performance on the test dataset.

## 8  Future Work

Although our method gets a competitive score, there are some drawbacks in our methods: the training time gets long with 85s/epoch, we can custom layers in the architecture to accelerate the computing cost. We can get more layers or ensemble more backbones to achieve higher results.
Another method we can approach by classifying each frame of image in the sequence of images, by applying transfer learning methods with the previous backbone, this method can achieve the higher score and reduce the overfitting problem with the small training dataset.